\documentclass[12pt,a4paper]{iopart}
\pdfoutput=1
\usepackage[T1]{fontenc}
\usepackage[latin9]{inputenc}
\usepackage[english]{babel}
\usepackage{color}
\usepackage{float}

\expandafter\let\csname equation*\endcsname=\relax
\expandafter\let\csname endequation*\endcsname=\relax

\usepackage{amsmath}
\usepackage{amssymb,amsthm}

\usepackage{graphicx}
\PassOptionsToPackage{normalem}{ulem}
\usepackage{ulem}
\usepackage{babel}

\makeatother

\begin{document}

\title[Phase Diagram of Twist Storing Lattice Polymers]{Phase Diagram of Twist Storing  Lattice Polymers in Variable Solvent Quality}
\author{E Dagrosa$^1$, A L Owczarek$^1$ and T Prellberg$^2$}
\address{$^1$ School of Mathematics and Statistics,
  The University of Melbourne, Parkville, Vic 3010, Australia.}
\address{$^2$ School of Mathematical Sciences, Queen Mary University
  of London, Mile End Road, London E1 4NS, UK.}
\ead{edagrosa@outlook.com, owczarek@unimelb.edu.au, t.prellberg@qmul.ac.uk}

\begin{abstract}
When double stranded DNA is turned in experiments it undergoes a transition.
We use an interacting self-avoiding walk on a three-dimensional fcc
lattice weighted by writhe to relate to these experiments and treat this problem via
simulations. We provide evidence for the existence of a thermodynamic phase transition induced by writhe
and examine related phase diagrams taking solvent quality and stretching into
account. 
\end{abstract}

\section{Introduction}

\begin{figure}[ht!]
\centering\includegraphics[width=6cm]{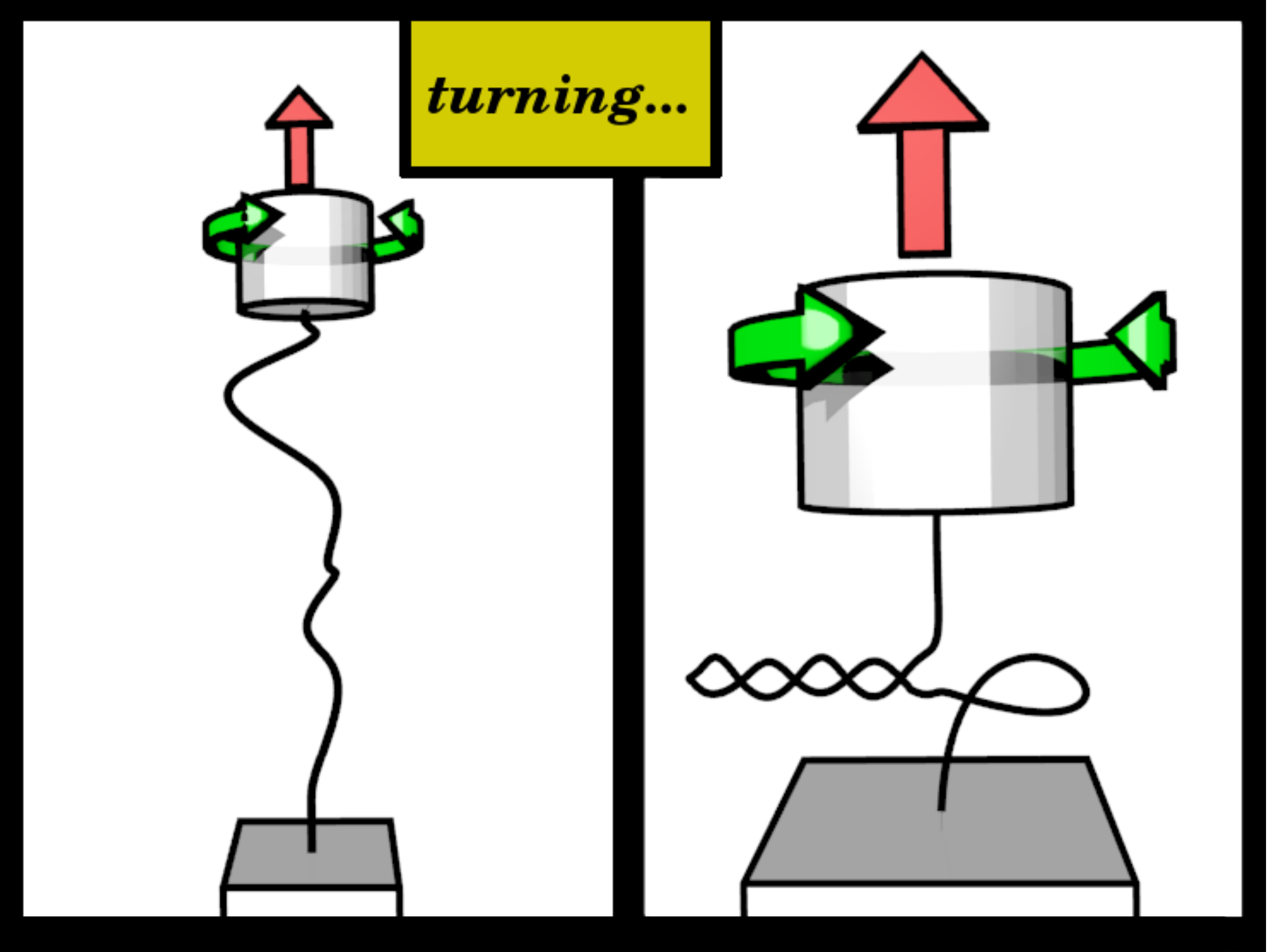}
\vspace{0cm}\protect\caption{\label{fig:FCCWALK_experimental_setup}Sketch of an experiment that
turns (green curved arrows) double stranded DNA under tension (red vertical arrow).
Initially, most of the turns are expected to be absorbed in the form
of twist. Due to the twist rigidity of the molecule, twisting causes
the DNA to exert a restoring torque on the cylinder. Upon turning
sufficiently, the DNA transitions into a regime where it is expected
to form plectonemes. }
\end{figure}

Since the mid 1990s, experiments on single molecules of double stranded
DNA have been performed \cite{Experiment_S007961070000018320000101}.
In some of these experiments, the molecule is held torsionally constrained
at a constant stretching force. Upon turning the molecule sufficiently,
the DNA transitions from a regime of stretched states into the supercoiled
regime. In this regime, the DNA is believed to coil around itself
to form plectonemes and the torque exerted on the apparatus plateaus
over the number of turns. Such an experiment is sketched in Figure~\ref{fig:FCCWALK_experimental_setup}.
In between the stretched and supercoiled regime lies the so called
buckled regime.

In 2008 \cite{ABRUPT_BUCKLING} and thereafter \cite{edsgcl.29795367720100101},
it was observed that, depending on the salinity of the solvent, the
transition between the regimes becomes rather abrupt. This phenomena
has been dubbed abrupt buckling transition.

In the literature, some of the mechanics of the experiments is modelled
by treating DNA as a mathematical ribbon. According to the Calegarenu-White-Fuller
theorem \cite{1969_WHITE_SL,CALUGAREANU59,CALUGAREANU_61_A,CALUGAREANU_61_B,Fuller_Decomp}, this allows a decomposition of the number of turns added to the
molecule, into writhe, total twist and a constant. More details on
mathematical modeling of DNA in general can be found in \cite{Fuller_Decomp,melb.b397887820090101}.
Details relating particular to the mechanics of the performed experiments,
and which are relevant for this paper, are given in \cite{cond-mat/990401819990401,Sinha_PhysRevE.85.041802,1304.356920130412}.

There are different approaches towards modeling such experiments.
Free energy models are very phenomenological but can provide exact
solutions. They allow the inclusion of many details related to the
DNA molecule. Specifically, in \cite{MULTI_PLECTONEME_PHASE,000299123300006n.d.,Marko_1997}
the free energy is developed around three stable solutions of the
elastic rod called straight, buckled and supercoiled (post-buckled).
However, these models do not include potential fluctuations that are
induced by correlations across the whole molecule. This makes them
mathematically less interesting. Also, although in these models the
buckling transition is set up to be a first-order phase transition,
it is arguable that this transition is really just an abrupt transition
(but not a phase transition) in a more fundamental model.

The alternative approach is to analyze the partition function directly.
To obtain useful results, a basic general model must incorporate several
features. Structures like plectonemes emerge only as a consequence
of self-avoidance of the polymer model and also writhe (and therefore
linking number) only exists for self-avoiding curves. Next, one
needs to be able to pick only states of a given linking number. Finally,
the ensemble needs to contain only configurations without knots. This
restriction to a knot type makes it impossible to date to treat the
problem analytically.

In this paper, we consider the second approach. However, we are not
interested in details specific to DNA. Instead, we consider the experiment
performed on the interacting self-avoiding walk (ISAW). The most famous
ISAW model weights SAWs by the number of contacts between distant vertices.
This model exhibits a second-order phase transition from a random
coil phase into a globule phase as the quality of the solvent decreases.
Variants of this model apply a pulling force which makes the collapse
transition first-order \cite{Lai_PhysRevE.58.6222}. Other models
consider an energy cost associated with bending the walk. When the
walk is sufficiently stiff, the collapse transition becomes a first
order crystallization transition from the coiled phase into the crystal
phase. Therefore, the phase diagram spanned by solvent quality and
stiffness contains a triple point with the three phases coil, crystal
and globule \cite{Grassberger_phasetransitions} incident. 

The ISAW model used here weights unknotted SAWs on the face centered
cubic lattice by their writhe. We will argue that this relates to
interpreting the SAW as a twist storing lattice polymer on which the
above experiment is performed. We use the Wang-Landau-Algorithm \cite{WANG_LANDAU_ORIG}
to analyze the model for scaling in the derivatives of the free energy.
This allows us to infer possible phase transitions.

When the knot type is not restricted, the corresponding model has
been treated by a mapping onto an $O\left(N\rightarrow0\right)$ field
theory \cite{Moroz1997695}. In this work it was suggested that a sufficiently
strong coupling to writhe collapses the walk. In a later work \cite{Kung}
a writhe induced first-order transition was predicted. In fact, we had 
previously performed corresponding simulations \cite{Dagrosa}
and found signs of a first-order transition, which turned out to be
a transition between knot types.

In this work, we find signs of a second-order transition upon weighting
walks by their writhe when knot type is fixed. Although, this transition is marked by fluctuations
in self-contacts, we argue that it is not the collapse transition,
but a transition into a writhed phase. In addition, we consider the
effect of stiffness and pulling to observe behaviour qualitatively
compatible with experiments performed on DNA.

\section{Self-Avoiding Walks as a Model of Twist Storing Polymers }
\begin{figure}[ht!]
\centering\includegraphics[width=8cm]{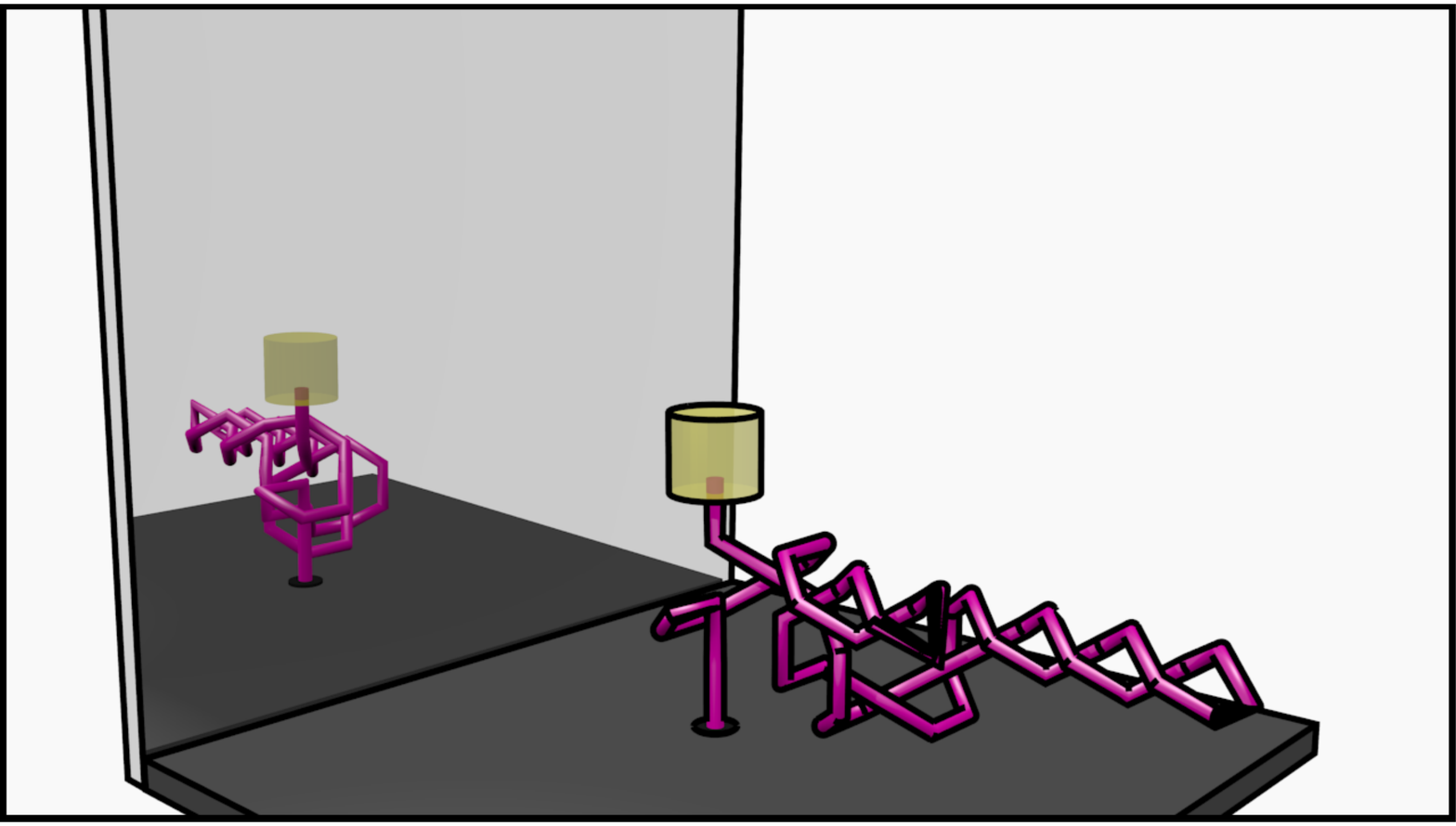}\caption{\label{fig:PULLED_SAW_FCC}A sketch of a SAW on the fcc lattice. The
SAW is attached to a surface and a cylinder mimicking conditions of
experiments that turn TSPs.}
\end{figure}
In order to model mechanics of turning a twist storing lattice polymer,
we require the definition of a lattice ribbon. For our purposes, we
think of a lattice ribbon $\mathcal{L}=\left(\varphi,\, V\right)$
as a pair formed by a self-avoiding walk (SAW) $\varphi$ and a vector
field $V$ on $\varphi$, so that we have two numbers $w$ and $t$
that are multiples of writhe and twist, respectively. In particular,
let $\varphi$ be a SAW on the non-normalized face centered cubic
lattice. We let $\varphi$ be geometrically constrained, so that the
first and last vertex lie on a given force axis and so that all vertices
of $\varphi$ lie in between planes through the first and last vertex.
We also restrict the knot type of $\varphi$ to be equivalent to the
unknot.

The following provides more details. Let $O=\left(0,\,0,\,0\right)$
be the origin of the face-centered cubic (fcc) lattice $\Lambda=\left\{ O+\sum_{\mathbf{k}}q_{k}\mathbf{k}|\, q_{k}\in\mathbb{Z},\,\mathbf{k}\in K\right\} $
where $K=\left\{ \left(1,\,1,\,0\right),\,\left(1,\,-1,\,0\right),...\right\} $.
A SAW $\varphi=\left\{ \varphi_{i}\right\} _{i=1,..,n+1}$ of length
$n$ consists of $n+1$ connected vertices $\varphi_{i}$ on $\Lambda$.
Therefore, with $\vec{\varphi}_{i}:=\varphi_{i+1}-\varphi_{i}$, $\vec{\varphi}_{i}^{2}=2$.
Also, let $\hat{\varphi}_{i}:=2^{-1/2}\vec{\varphi}_{i}$. We anchor
$\varphi$ at the origin, i.e. $\varphi_{1}=O$ and define a force
axis $\mathbf{f}:=\left(1,1,0\right)$. We require all vertices of
the walk to lie on or above a plane that lies perpendicular to $\mathbf{f}$
through $O$, i.e $\left(\varphi_{i}\right)_{x}+\left(\varphi_{i}\right)_{y}\geq0$.
The last vertex, $\varphi_{n+1}$ is required to lie on the axis $\varphi_{n+1}\overset{!}{=}O+(h-2)/2\,\mathbf{f}$,
for some integer $h/2$. We then append another vertex $\varphi^{(final)}=\varphi_{n+1}+\mathbf{f}$
to the walk. This vertex has no degree of freedom so it does not count
towards the length, however we require it to properly define the writhe
of this non-closed walk. We imagine a second plane through $\varphi^{(f)}$
such that no vertex can lie on or above this plane, i.e. $\left(\varphi_{i}\right)_{x}+\left(\varphi_{i}\right)_{y}<2\left(\varphi^{(final)}\right)_{x}$.
Finally, we imagine to connect $\varphi_{1}-\mathbf{f}$ and $\varphi_{n+1}$
through a planar walk $\nu$ whose arch is wide enough so that its
rotation ellipsoid around $\mathbf{f}$ is disjoint with $\varphi$.
Then, we require that the closed polygon $\bar{\varphi}:=\varphi\cup\nu$
is an unknot. Formally, the writhe of $\varphi$ can be defined by
the writhe of $\bar{\varphi}$, i.e. $Wr\left(\varphi\right):=Wr\left(\bar{\varphi}\right)$.
To compute the writhe, we use the formula
\begin{eqnarray}
Wr\left(\varphi\right) & = & Lk\left(\bar{\varphi},\,\bar{\varphi}+\epsilon\mathbf{d}\right)-\frac{1}{2\pi}\sum_{i=1}^{n}\Biggl\{\label{eq:WR_polygonal_writhe}\\
 &  & \arctan\left(\frac{\left\langle \mathbf{d},\,\hat{\varphi}_{i-1}\right\rangle -\left\langle \mathbf{d},\,\hat{\varphi}_{i}\right\rangle \,\left\langle \hat{\varphi}_{i-1},\,\hat{\varphi}_{i}\right\rangle }{\left[\hat{\varphi}_{i-1},\,\hat{\varphi}_{i},\,\mathbf{d}\right]}\right)\nonumber \\
 &  & +\arctan\left(\frac{\left\langle \mathbf{d},\,\hat{\varphi}_{i}\right\rangle -\left\langle \mathbf{d},\,\hat{\varphi}_{i-1}\right\rangle \,\left\langle \hat{\varphi}_{i-1},\,\hat{\varphi}_{i}\right\rangle }{\left[\hat{\varphi}_{i-1},\,\hat{\varphi}_{i},\,\mathbf{d}\right]}\right)\Biggr\}\nonumber,
\end{eqnarray}
where $Lk\left(\bar{\varphi},\,\bar{\varphi}+\epsilon\mathbf{d}\right)$
is the linking number between the closed polygon $\bar{\varphi}$
and a copy of $\bar{\varphi}$ pushed off a small amount into direction
$\mathbf{d}$. We set $\hat{\varphi}_{0}:=\hat{\mathbf{f}}$. $\left\langle \bullet,\bullet\right\rangle $
denotes the scalar product, and $\left[\bullet,\,\bullet,\,\bullet\right]$
the triple product. Using an appropriate algorithm to compute $Lk\left(\bar{\varphi},\,\bar{\varphi}+\epsilon\mathbf{d}\right)$
and a proper choice for $\mathbf{d}$ , one may show that writhe can
be computed without taking the closure $\nu$ of $\bar{\varphi}$
into account. We have derived the formula~\ref{eq:WR_polygonal_writhe}
in \cite{GENERLARIBBON}. Alternative expressions for the writhe of
polygonal curves can be found in \cite{SUMNERS_WRITHES,Klenin,aldinger_formulae},
however some of these are not appropriate for the kind of simulation
performed here. For reasons explained in the next section, we use
the observables $w=30\, Wr\left(\varphi\right)$, $t=30\, Tw\left(V,\,\varphi\right)$
and $l=w+t$.

Furthermore, let $h=2\,\left(\varphi^{(final)}\right)_{x}$ measure
$\sqrt{2}$ times the distance of the final vertex from $O$. Let
$c$ denote the number of contacts between vertices and let $s=n+\sum_{i}\left\langle \vec{\varphi}_{i},\,\vec{\varphi}_{i+1}\right\rangle $
denote the number of stiff sites. Note that $s\in\left[0,\,3n\right]$.
We hold the temperature $T$ fixed and define the energy of a configuration
$\mathcal{L}$ by 
\[
E\left(\mathcal{L}\right)=\mbox{const}-k_{B}T\,\left(\beta_{h}h+\beta_{c}c+\beta_{s}\, s+\beta_{t}\, t\right),
\]
where the coupling $\beta_{h}$ is related to pulling force, $\beta_{s}$
to stiffness (bending energy), $\beta_{t}$ to twist rigidity and
$\beta_{c}$ can be related to the quality of the solvent. We remark
that the subscripts to the betas are merely a pairing label, not an
index, i.e., $\beta_{c}$ does not depend on $c$. To keep notation
short, we collect some of the observables as $\mathbf{m}=\left(s,\, h\, c\right)^{T}$
and the corresponding couplings as $\vec{\beta}=\left(\beta_{s},\,\beta_{h},\,\beta_{c}\right)^{T}$.
We denote the density of states by $C_{n\,\mathbf{m}\, w\, t}^{\left(Rib\right)}$,
so that the partition function of the constant linking number $l$
ensemble is given by 
\[
Z_{n\, l}^{\left(Rib\right)}\left(\beta_{t},\,\vec{\beta}\right)=\mbox{const}\,\sum_{w}\sum_{t}\sum_{\mathbf{m}}C_{n\,\mathbf{m}\, w\, t}^{\left(Rib\right)}\, e^{\vec{\beta}\cdot\mathbf{m}}e^{\beta_{t}t}\delta_{l,\, w+t},
\]
where, using the CFW theorem, $\delta_{l,\, w+t}$ picks only the ribbons
that have linking number $l$. Discrete Laplace transformation of
$Z_{n\, l}$ with respect to $l$ yields the conjugated fully canonical
ensemble of ribbons
\begin{eqnarray*}
Z_{n}^{\left(Rib\right)}\left(\beta_{t},\,\vec{\beta},\,\beta_{l}\right) & = & \sum_{l}\, Z_{n\, l}^{\left(Rib\right)}\left(\beta_{t},\,\vec{\beta}\right)\, e^{\beta_{l}l}\\
 & = & const\,\sum_{w}\sum_{t}\sum_{\mathbf{m}}C_{n\,\mathbf{m}\, w\, t}^{\left(Rib\right)}\, e^{\vec{\beta}\cdot\mathbf{m}}e^{\beta_{l}w}e^{\left(\beta_{t}+\beta_{l}\right)t},
\end{eqnarray*}
where $\beta_{l}$ is related to torque. This ensemble corresponds
to an experiment that keeps the polymers under a fixed torque rather
than linking number. Therefore, we refer to it as the constant torque
ensemble. So far, we have not specified how the vector field $V$
should look. Suppose that a rule of constructing $V$ is such
that the number of ribbon frames with total twist $t$, i.e. $C_{t}$
is independent of $\varphi$, then 
\begin{equation}
C_{n\,\mathbf{m}\, w\, t}^{\left(Rib\right)}=C_{n\,\mathbf{m}\, w}C_{t}\label{eq:Fac},
\end{equation}
and the partition function factors according to 
\begin{equation}
Z_{n}^{\left(Rib\right)}\left(\beta_{t},\,\vec{\beta},\,\beta_{l}\right)=Z_{n}^{\left(Wr\right)}\left(\vec{\beta},\,\beta_{l}\right)Z_{n}^{\left(Tw\right)}\left(\beta_{t},\,\beta_{l}\right)\label{eq:factZ},
\end{equation}
where 
\begin{equation}
Z_{n}^{\left(Wr\right)}\left(\vec{\beta},\,\beta_{l}\right)=\sum_{w}\sum_{\mathbf{m}}C_{n\,\mathbf{m}\, w}\, e^{\vec{\beta}\cdot\mathbf{m}}e^{\beta_{l}w}\label{eq:Z_WR}
\end{equation}
and $Z_{n}^{\left(Tw\right)}\left(\beta_{t},\,\beta_{l}\right)=\mbox{const}\,\sum_{t}C_{t}e^{\left(\beta_{t}+\beta_{l}\right)t}$.
As we are interested in the singularity structure of the free energy
$f^{\left(Rib\right)}=\lim_{n\rightarrow\infty}f_{n}^{\left(Rib\right)}$,
the factoring~(\ref{eq:factZ}) of the partition function implies
that we may examine the scaling of the derivatives of $f_{n}^{\left(Wr\right)}\left(\beta_{t},\,\vec{\beta},\,\beta_{l}\right)=n^{-1}\log Z_{n}^{\left(Wr\right)}$
and $f_{n}^{\left(Tw\right)}$ independently to obtain information
of the potential critical points of the system. However, because total
twist is a quantity that is computed along the walk, we do not expect
any singularities associated with $f^{\left(Tw\right)}$. Consequently,
we can focus on~(\ref{eq:Z_WR}), which only requires us to perform
simulations of a SAW rather than a ribbon. We note that even in the
case $\vec{\beta}=0$, our partition function $Z_{n}^{\left(Wr\right)}$
contains a Boltzmann factor $\exp\left(\beta_{l}\, w\right)$, where
the writhe $w$ is obtained by relating near and distant parts of
the walk to each other. Therefore, the torque $\beta_{l}$ can induce
long distance correlations across the polymer. But this means that
the system might potentially become critical. In fact, we have examined
the partition function $\sum C_{n\, w}e^{\beta_{l}w}$ in systems
which do not preserve the topology \cite{Dagrosa}. In this case,
we found a first-order transition between knot types associated with
the partition function.

We recall that the convenient factorization~(\ref{eq:factZ}) is
due to two reasons. First, we work in the conjugated constant torque
ensemble. In the thermodynamic limit $n\rightarrow\infty$, the ensembles
are equivalent so that any inference we make about phase transitions should
also hold in the linking number ensemble. The second reason is relation~(\ref{eq:Fac}).
Although, we can certainly make a rule so that (\ref{eq:Fac}) holds,
it is not necessarily natural on the fcc lattice. However, the resulting
factoring of the partition function~(\ref{eq:factZ}) is also obtained
in continuous models of dsDNA. For example, compare (\ref{eq:factZ})
to formula (6) in \cite{Sinha_PhysRevE.85.041802}, where the ribbon
frame is attached to the continuous worm like chain model. Therefore,
imposing condition~(\ref{eq:Fac}) yields a simple model compatible
with non-lattice models in the literature.

\section{Algorithm and Data}

\subsection{Algorithm}

Let $t$ be the algorithm time, then, at given length $n$, the Wang-Landau
Algorithm (WLA) can be used to obtain an estimate $s_{n\, w\,\mathbf{m}}^{(est)}\left(t\right)$
of the microcanonical entropy 
\[
s_{n\, w\,\mathbf{m}}:=\log C_{n\, w\,\mathbf{m}}\,,
\]
where one expects that for large enough times $s_{n\, w\,\mathbf{m}}^{(est)}\left(t\right)\approx const\, t+s_{n\, w\,\mathbf{m}}$.
We can consider the WLA as a pair $\left(\varphi\left(t\right),\, s_{n\, w\,\mathbf{m}}^{(est)}\left(t\right)\right)$,
where $\varphi\left(t\right)$ is the current state of the algorithm.
A step of the WLA algorithm consists in proposing a state $\varphi^{*}$
to the WLA, which is accepted with the probability
\begin{equation}
p_{acc}=\min\left[1,\,\exp\left\{ s_{w\left(t\right)\,\mathbf{m}\left(t\right)}^{est}\left(t\right)-s_{w^{*}\mathbf{m}^{*}}^{est}\left(t\right)\right\} \right]\label{eq:p_acc},
\end{equation}
where $s_{w\left(t\right)\,\mathbf{m}\left(t\right)}^{est}\left(t\right)$
is the value of $s_{n\, w\,\mathbf{m}}^{(est)}\left(t\right)$ at
$w=w\left(t\right)$ and $\mathbf{m}=\mathbf{m}\left(t\right)$. Therefore,
\[
\varphi\left(t+1\right)=\begin{cases}
\varphi^{*} & p\leq p_{acc},\\
\varphi\left(t\right) & else.
\end{cases}
\]
Finally, $s_{n\, w\,\mathbf{m}}^{(est)}\left(t\right)$ is updated
as $s_{n\, w\,\mathbf{m}}^{(est)}\left(t+1\right)=s_{n\, w\,\mathbf{m}}^{(est)}\left(t\right)+\log f\,\delta_{w\, w\left(t+1\right)}\delta_{\mathbf{m}\,\mathbf{m}\left(t+1\right)}$,
where $f$ is called the modification factor. For large $t$ the modification
factor is reduced as $\log f\sim1/t$. Instead of using the traditional
method we will use a modification of the WLA to obtain estimates for
\begin{equation}
s_{n\, w}\left(\vec{\beta}\right):=\log\left\{ \sum_{\mathbf{m}}C_{n\, w\,\mathbf{m}}e^{\vec{\beta}\cdot\mathbf{m}}\right\}.
\end{equation}
These are obtained by changing the acceptance probability to
\begin{equation}
p_{acc}=\min\left[1,\,\exp\left\{ s_{w\left(t\right)}^{est}\left(t\right)-s_{w^{*}}^{est}\left(t\right)+\vec{\beta}\cdot\left(\mathbf{m}^{*}-\mathbf{m}\left(t\right)\right)\right\} \right].
\end{equation}
One obtains $\varphi^{*}$ from $\varphi\left(t\right)$ by applying
a move on $\varphi\left(t\right)$. We use three kinds of move:
\begin{description}
\item [{Bond~Flip}] A vertex $\varphi_{i}$ is selected randomly. It is
randomly moved onto one of the lattice sites that neighbor both $\varphi_{i-1}$
and $\varphi_{i+1}$. 
\item [{Kink~Transport}] A vertex $\varphi_{i}$ is selected randomly.
If $\varphi_{i-1}$ and $\varphi_{i+1}$ are neighbors $\varphi_{i}$
is removed and inserted either between two different vertices or it
is appended at the end in $\hat{\mathbf{f}}$ direction. 
\item [{End~to~Kink}] When the last step of the walk lies in $\hat{\mathbf{f}}$
direction, the last vertex can be removed and inserted between two
vertices of the SAW. 
\end{description}
We note that every move has its inverse move. The moves are applied
randomly, but in a way that detailed balance is satisfied. When a
move yields a state that is not in the ensemble, for example the state
is not self-avoiding, we set $\varphi^{*}=\varphi\left(t\right)$.
The authors in \cite{2258711620120301} assumed that the combination of bond-flips and kink-transport (pull moves for a SAP) are ergodic within the knot type of a polygon on the sc lattice. We will assume the same to be true for the  restricted walk on the fcc lattice considered here. One can easily
convince oneself that the move set is ergodic within the desired ensemble
for length $n\leq4$. In dealing with the WLA, it is common practice
to split the space of macrostates $\Gamma=\left\{ w_{1},w_{2},\,...,w_{max}\right\} $
into overlapping subsets $\Gamma_{i}$ where $\Gamma=\bigcup_{i}\Gamma_{i}$
and $\Gamma_{i}\cap\Gamma_{i+1}\neq\emptyset$ and generate estimates
$s_{n\, w}^{\left(i\right)}\left(\vec{\beta},\, t^{\left(i\right)}\right)$
over $\Gamma_{i}$. This requires restricting $\varphi^{\left(i\right)}\left(t^{\left(i\right)}\right)$
so that $w^{\left(i\right)}\left(t^{\left(i\right)}\right)\in\Gamma_{i}$.
However, particularly for some $\vec{\beta}$ far away from $0$,
this method works very poorly and it becomes essential to parallelize
the WLAs $WLA^{\left(i\right)}\left(t^{\left(i\right)}\right)=\left(\varphi^{\left(i\right)}\left(t^{\left(i\right)}\right),\, s_{n\, w}^{(i)}\left(\vec{\beta},\, t^{\left(i\right)}\right)\right)$
by introducing an additional move. Consider two pairs of Wang-Landau
algorithms $WLA^{\left(i\right)}\left(t^{\left(i\right)}\right)$
and $WLA^{\left(j\right)}\left(t^{\left(j\right)}\right)$. Denote
by $t$ a global time, so that at this time the corresponding chains
are in the states $\varphi^{\left(i\right)}\left(t^{\left(i\right)}\left(t\right)\right)$
and $\varphi^{\left(j\right)}\left(t^{\left(j\right)}\left(t\right)\right)$
respectively. For now, denote by $w^{\left(i\right)}\left(t\right)$
the writhe of $\varphi^{\left(i\right)}\left(t^{\left(i\right)}\left(t\right)\right)$
and by $s_{w^{\left(j\right)}}^{\left(i\right)}$ the current estimate
of $s_{w}^{\left(i\right)}$ at $w=w^{\left(j\right)}\left(t\right)$.
Then, when the algorithms $WLA^{\left(i\right)}$ and $WLA^{\left(j\right)}$
overlap at global time $t$, i.e. $w^{\left(i\right)}\left(t\right),\, w^{\left(j\right)}\left(t\right)\in\Gamma^{\left(i\right)}\cap\Gamma^{\left(j\right)}$
a chain switch is performed with probability
\begin{equation}
p_{acc}=\min\left[1,\,\exp\left\{ s_{w^{\left(i\right)}}^{\left(i\right)}-s_{w^{\left(j\right)}}^{\left(i\right)}+s_{w^{\left(j\right)}}^{\left(j\right)}-s_{w^{\left(i\right)}}^{\left(j\right)}\right\} \right],
\end{equation}
so that 
\begin{equation}
\varphi^{\left(i\right)}\left(t^{\left(i\right)}+1\right)=\begin{cases}
\varphi^{\left(j\right)}\left(t^{\left(j\right)}\right) & p\leq p_{acc},\\
\varphi^{\left(i\right)}\left(t^{\left(i\right)}\right) & else.
\end{cases}
\end{equation}
The details of this procedure were suggested in \cite{PARADIGM_WL},
where it was shown to be very helpful on some benchmark problems.
In our case, it is essential to overcome the trapping of the walk
$\omega^{\left(i\right)}$ in regions of the state space that are
not representative enough to produce a good estimate of the entropy.

\subsection{Data}

\begin{figure}[ht!]
\centering\includegraphics[width=7cm]{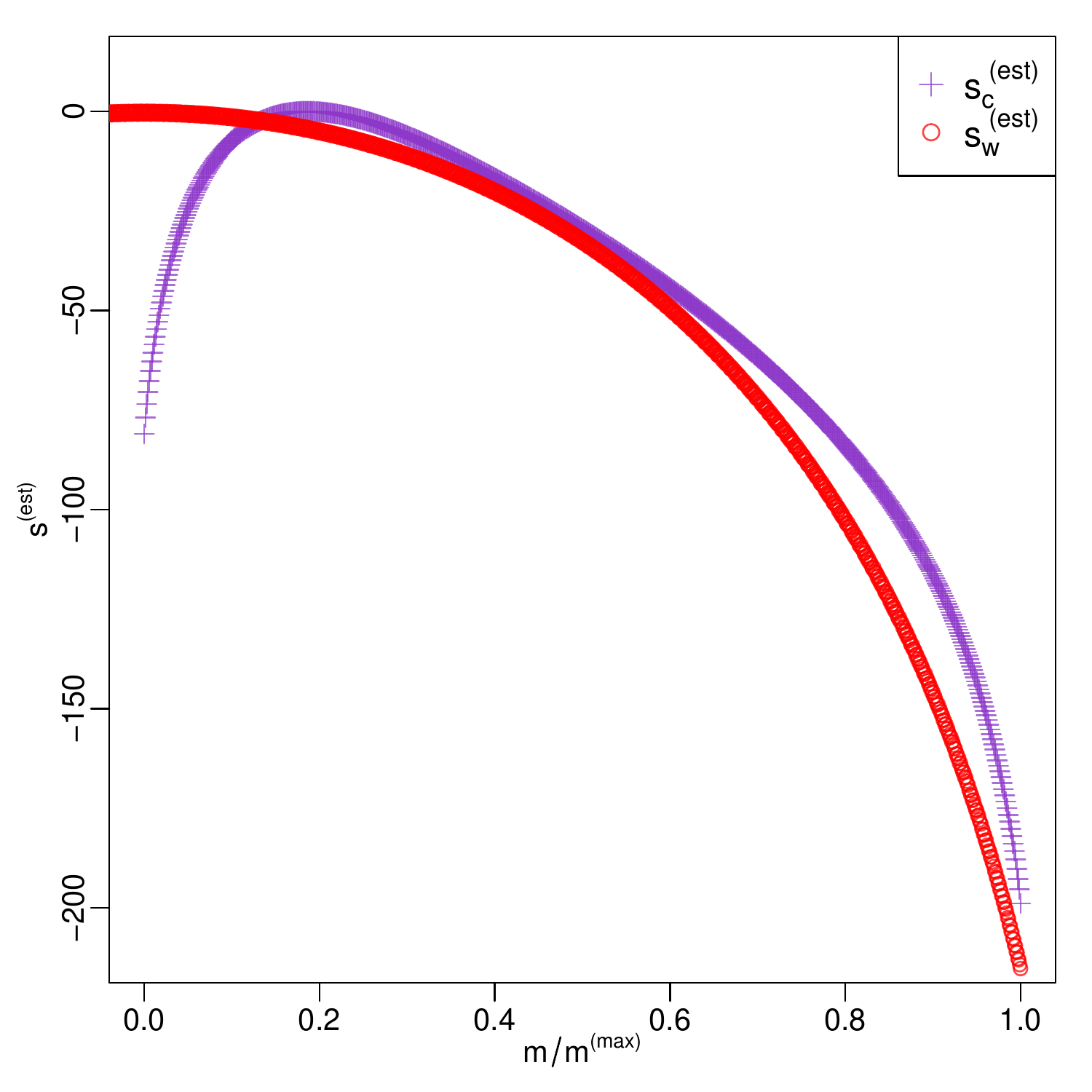}

\protect\caption{\label{fig:Entropies}Shown are entropies over $c/715$ and over $w/1020$ at $n=192$. 
We note that $w^{\left(max\right)}=1020$
is the value of the cut off and not the actual maximal value of the
writhe $w$. Using the asymptotic formula \cite{Thesis} $\lim_{n\rightarrow\infty}Wr^{max}/n=\frac{1}{2\pi}\arctan\left(2\sqrt{2}\right)$
for the maximum writhe of an unknot on the fcc lattice, we expect
the maximum $w$ near 1128. }
\end{figure}

We ran several simulations to obtain estimates for $s_{n\, w}\left(\beta_{s},\,\beta_{h},\,\beta_{c}\right)$
and \newline $s_{n\, c}\left(\beta_{l},\,\beta_{s},\,\beta_{h}\right)$.
Although writhe assumes discrete values on the fcc lattice, the macrostates labeled
by $w$ lie very dense, thus the estimates of $s_{n\, w}$ are made
using $w\leftarrow round(abs(w))$ and taking advantage of the symmetry
$s_{n\, w}=s_{n\,-w}$. Recall that $w:=30\, Wr$, so that $30$ serves
as the resolution for the estimates of $s_{n\, w}$. This resolution
factor was determined, as $round(abs(30\, Wr))$ allows to just resolve
the state of maximum writhe of self-avoiding polygons on the fcc lattice.
As an example of the results, Figure~\ref{fig:Entropies} shows
the estimated entropies $s_{n=192\, w}\left(\beta_{s}=0,\,\beta_{h}=0,\,\beta_{c}=0\right)$
and $s_{n=192\, c}\left(\beta_{l}=0,\,\beta_{s}=0,\,\beta_{h}=0\right)$.

As we are interested in the potentially critical points in coupling
space $\left(\beta_{l},\,\beta_{s},\,\beta_{h},\,\beta_{c}\right)$
we want to observe the scaling of the derivatives of the free energy
$f_{n}^{\left(Wr\right)}$. With $m=l,\, c,\, s,\,...$ define
\begin{equation}
f_{n}^{m_{1},\, m_{2}...,\, m_{k}}\left(\beta_{l},\,\vec{\beta}\right):=\left(\frac{\partial^{k}}{\partial\beta_{m_{1}}\partial\beta_{m_{2}}...\partial\beta_{m_{k}}}f_{n}^{\left(Wr\right)}\left(\beta_{l},\,\vec{\beta}\right)\right)^{(est)},
\end{equation}
where for example $f_{n}^{l}=n^{-1}\left\langle w\right\rangle _{n}^{\left(est\right)}\left(\beta_{l},\,\vec{\beta}\right)$,
which can be computed as 
\begin{equation}
\left\langle w\right\rangle _{n}^{\left(est\right)}\left(\beta_{l},\,\vec{\beta}\right)=\frac{\sum_{w}w\, e^{\beta_{l}\, w+s_{n\, w}^{\left(est\right)}\left(\vec{\beta}\right)}}{\sum_{w}e^{\beta_{l}\, w+s_{n\, w}^{\left(est\right)}\left(\vec{\beta}\right)}}.
\end{equation}
In the same way one uses $s_{n\, w}^{(est)}\left(\vec{\beta}\right)$
to compute all derivatives with respect to $\beta_{l}$, i.e. $f_{n}^{l},f_{n}^{l\,l},\ldots$. To
obtain the estimates $f_{n}^{m}$,~$f_{n}^{m\, m}$ with $m\neq l$,
note that for a generic function $g\left(w,\,\mathbf{m}\right)$
\begin{eqnarray}
\left\langle g\left(w,\,\mathbf{m}\right)\right\rangle _{n}\left(\beta_{l},\,\vec{\beta}\right) & := & \frac{\sum_{w}e^{\beta_{l}w}\sum_{\mathbf{m}}g\left(w,\,\mathbf{m}\right)e^{\vec{\beta}\cdot\mathbf{m}+s_{n\, w\,\mathbf{m}}}}{\sum_{w}e^{\beta_{l}w}\sum_{\mathbf{m}}e^{\vec{\beta}\cdot\mathbf{m}+s_{n\, w\,\mathbf{m}}}}\nonumber \\
 & = & \frac{\sum_{w}e^{\beta_{l}w}\left\langle g\left(w,\,\mathbf{m}\right)\right\rangle _{w}e^{s_{n\, w}\left(\vec{\beta}\right)}}{\sum_{w}e^{\beta_{l}w+s_{n\, w}\left(\vec{\beta}\right)}},
\end{eqnarray}
where $\left\langle g\left(w,\,\mathbf{m}\right)\right\rangle _{w}=\sum_{\mathbf{m}}g\left(w,\,\mathbf{m}\right)e^{\vec{\beta}\cdot\mathbf{m}+s_{n\, w\,\mathbf{m}}-s_{n\, w}\left(\vec{\beta}\right)}$.
We obtain estimates for $\left\langle m\right\rangle _{w}$ as the
sample averages $\left\langle m\right\rangle _{w}^{\left(est\right)}:=N_{Samples}^{-1}\sum_{i=1}^{N_{Samples}}m\left(w\left(t_{i}\right)=w\right)$.
Therefore, every time $\varphi\left(t\right)$ has writhe $w$, we
take a sample $m\left(\varphi\left(t_{i}\right)\right)$. In the same
way we obtain estimates of $\left\langle m^{2}\right\rangle _{w}^{\left(est\right)}$.

When the states $\varphi^{*}$ proposed to the algorithm are drawn
according to the probability distribution the form of the error is
well known and statistical \cite{Understanding_Wang_Landau}. However,
in our case, the states $\varphi^{*}$ are proposed by applying a
move to $\varphi\left(t\right)$. This is usually associated with
an unknown systematic error related to the move set. Yet, one hopes
that that the error decays with the modification factor $\log f\sim1/t$.
We also expect some residual boundary effect error due to the splitting
of $\Gamma$ into $\Gamma_{i}$. Finally, we expect a statistical
error associated with combining the $s_{n\, w}^{\left(i\right)}\left(t^{\left(i\right)}\right)$
to form $s_{n\, w}^{\left(est\right)}$. In conclusion, we can not
predict the form of the error, so we form a naive standard error.
Therefore we provide standard confidence intervals for $f_{n}^{l\, l}\left(\beta_{l},\,\vec{\beta}\right)$
by averaging over several $s_{n\, w}^{\left(est\right)}$. These estimates
are obtained at different global times, but with small enough modification
factor. Typically, it is required that $\log\left(f\right)<10^{-8}$.

\section{Results}

\subsection{Flexible polymers without pulling}
\begin{figure}[ht!]
\centering\includegraphics[width=10cm]{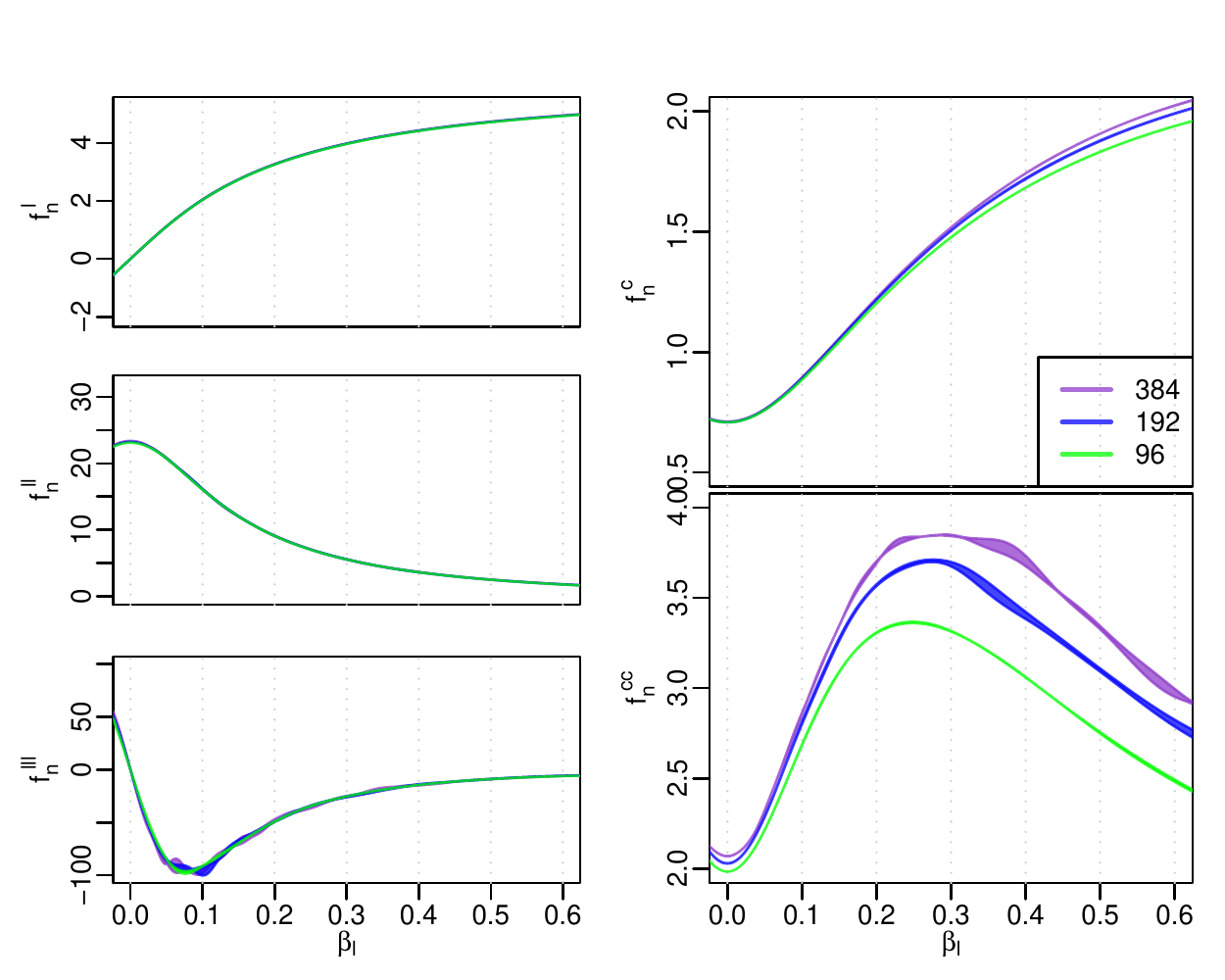}

\caption{\label{fig:w_000}Scaling of the first three derivatives of the free
energy with respect to $\beta_{l}$ against $\beta_{l}$ and the first
two derivatives with respect to $\beta_{c}$. The curves of $f_{n}^{l}$,
$f_{n}^{ll}$ and $f_{n}^{lll}$ are almost identical at the considered
lengths $384,\,192$ and $96$. }
\end{figure}

First we consider the case of flexible lattice polymer that are not
being pulled ($\beta_{s}=0$, $\beta_{h}=0$). The left-hand side
graphs in Figure~\ref{fig:w_000} show the scaling of the first,
second and third derivatives of the free energy with respect to $\beta_{l}$
against the torque $\beta_{l}$. None of the derivatives shows any
signs of non-trivial scaling. The graphs on the right-hand side in
Figure~\ref{fig:w_000} show the estimates for the first and second
derivative with respect to $\beta_{c}$. The graphs of $f_{n}^{cc}\left(\beta_{l}\right)$
have a peak between $\beta_{l}=0.2$ and $\beta_{l}=0.4$. The height
of the peak $f_{n}^{cc}\left(\beta_{l}\right)$ increases with $n$.
However, the scaling is somewhat unusual because the scaling region
extends far to the right of the peak position. Nevertheless, we conjecture
that this scaling is associated with a continuous phase transition,
for which we associate the position of the peak with the location
of the transition. In the following we will provide additional evidence
for this transition by considering the problem in the $\beta_{c}$-$\beta_{l}$
plane.

\begin{figure}[ht!]
\centering\includegraphics[width=8cm]{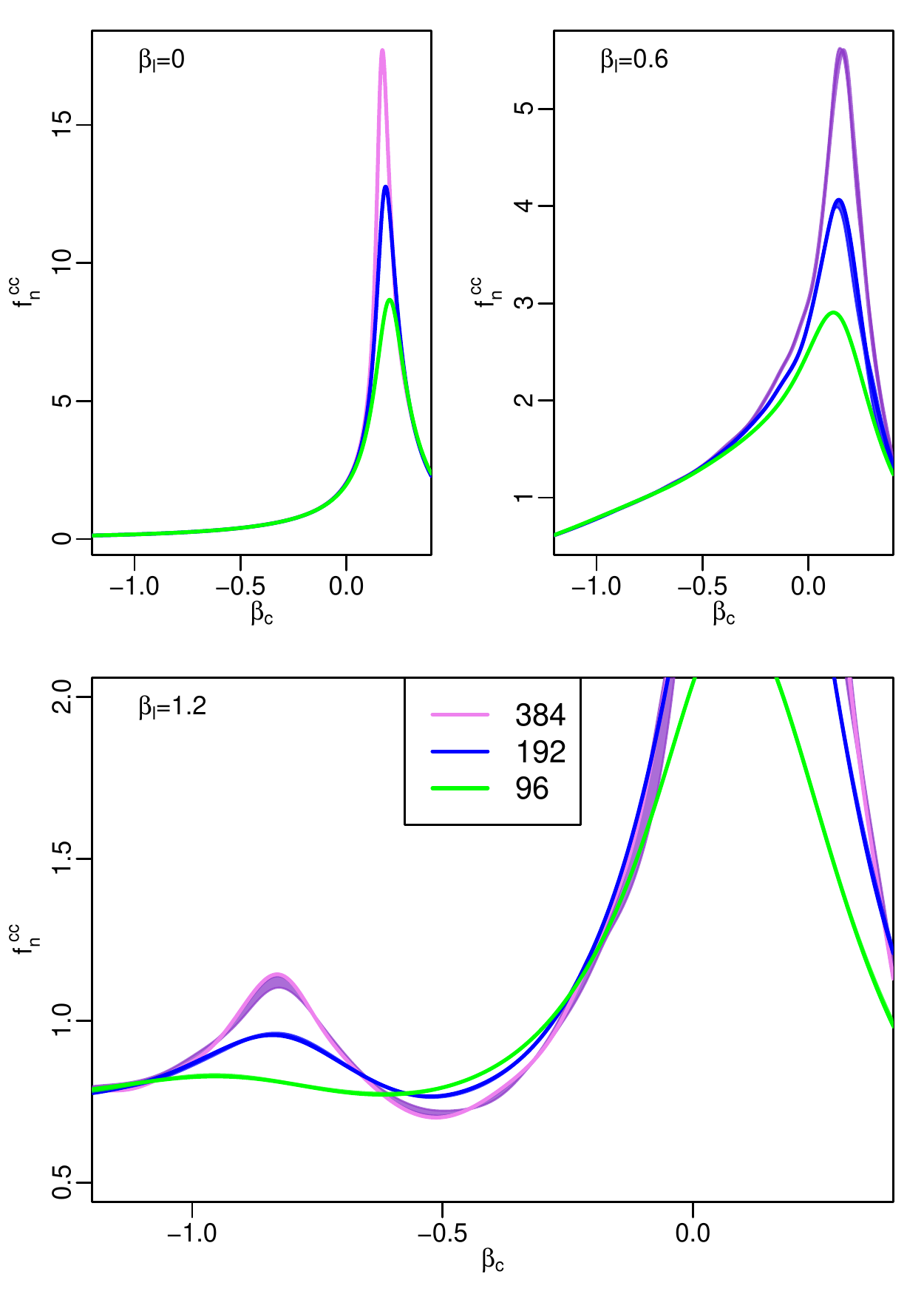}

\caption{\label{fig:c_00_scaling}Scaling of $f_{n}^{cc}\left(\beta_{l}=\mbox{const},\,\beta_{c},\,\beta_{s}=0,\,\beta_{h}=0\right)$
at different torques $\beta_{l}=\left\{ 0,\,0.6,\,1.2\right\} $. }
\end{figure}

\begin{figure}[ht!]
\centering\includegraphics[width=10cm]{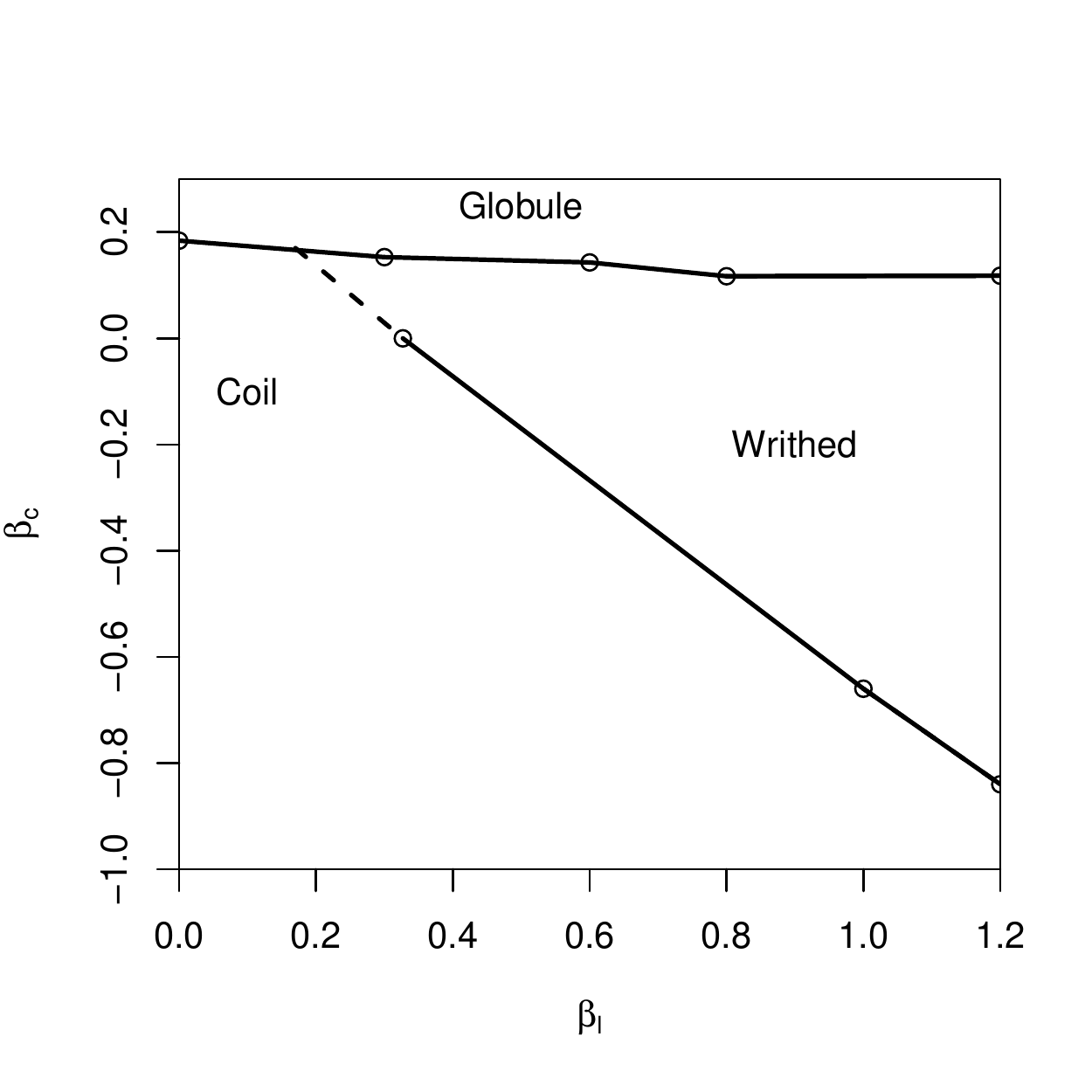}

\caption{\label{fig:w_000-1} Finite size schematic phase diagram at $\beta_{s}=0$, $\beta_{h}=0$.
The circles correspond to the positions of peaks in $f_{n=192}^{cc}$.
The solid lines are linear interpolation between the circles and represent
the phase boundaries. The dotted line is a linear interpolation of the
coil-writhed phase boundary. It meets the coil-globule transition
at the conjectured triple point near $\left(\beta_{l}=0.17,\,\beta_{c}=0.17\right)$.
No line in this diagram is expected to correspond to a first-order transition.}
\end{figure}

Figure~\ref{fig:c_00_scaling} shows the scaling $f_{n}^{cc}\left(\beta_{c},\,\beta_{l}=\mbox{const}\right)$
at different torques. At $\beta_{l}=0$ one observes scaling around
$\beta_{c}=0.2$, which is compatible with a second-order phase transition.
The transition can be identified with the familiar collapse transition
from the coil into the globule phase. As the torque is increased a
transition into the globule phase persists around $\beta_{c}=0.2$.
In addition to this transition, Figure~\ref{fig:c_00_scaling}
shows that at $\beta_{l}=1.2$, scaling around $\beta_{c}=-0.8$ emerges.
This scaling is potentially compatible with an additional continuous
phase transition. At the considered lengths, we do not find signs
of this latter transition in $f_{n}^{cc}\left(\beta_{c},\,\beta_{l}=\mbox{const}\right)$
when $\beta_{l}\leq0.8$. However, we conjecture that this transition
is related to the peak in $f_{n}^{cc}\left(\beta_{l},\,\beta_{c}=0\right)$
to form the finite size phase diagram sketched in Figure~\ref{fig:w_000-1}. 

We estimate  a triple point near $\left(\beta_{l}=0.17,\,\beta_{c}=0.17\right)$
where the three phases coil, globule and writhed are incident. The writhed phase
is the conjectured phase, which encompasses the states of intermediate
and high writhe but which are not part of the globule phase. One notes
that all transitions: coil-globule, coil-writhed and writhed-globule
are associated with contact fluctuations. When the transitions are
close in the space of couplings, we can imagine that the associated
fluctuations overlap. Therefore it is not possible to locate the coil-writhed
transition, as it is overpowered by the fluctuations associated with
the writhed-globule transition. It is known that the contact fluctuations
diverge logarithmically across the collapse transition. If this is
also true for the writhed-globule transition, and assuming that the
fluctuations associated with the coil-writhed transition do not diverge,
then in general we can not expect to resolve the coil-writhed transition
in $f_{n}^{cc}\left(\beta_{c},\,\beta_{l}=\mbox{const}\right)$ at any length
when the writhed-globule transition lies close. 

Consider the effect of negative $\beta_{c}$. First, negative $\beta_{c}$
corresponds to a repulsion between neighboring monomers and thus a
swelling of the chain. However, the repulsion is short range, that
is one lattice unit. Therefore, as the length is increased, this effect
becomes negligible compared to the swelling due to the entropy. 

The second effect of negative $\beta_{c}$ is the suppression of $2\,\pi/3$
angles between bonds. These correspond to kinks, i.e. local configurations
where two next nearest vertices of the SAW are one lattice step apart.
Thus, negative $\beta_{c}$ forces a certain type of stiffness on
the lattice polymer. As discussed above, stiffness has a non-trivial
effect on the collapse transition. We also expect stiffness to be
relevant to writhe. Therefore, we will consider stiff (semi-flexible)
polymers in the following. 

\subsection{Semi-flexible pulled polymers}
We now consider the case of semi-flexible polymers ($\beta_{s}>0$)
being subject to a strong pulling force. This makes it
is easier to converge the estimates $s_{n\, w}^{\left(est\right)}\left(\vec{\beta}\right)$
(and in particular $s_{n\, c}^{\left(est\right)}\left(\vec{\beta}\right)$)
properly. Also, the case of semi-flexible, pulled polymers corresponds
to the situation in experiments on dsDNA.

\begin{figure}[ht!]
\centering\includegraphics[width=10cm]{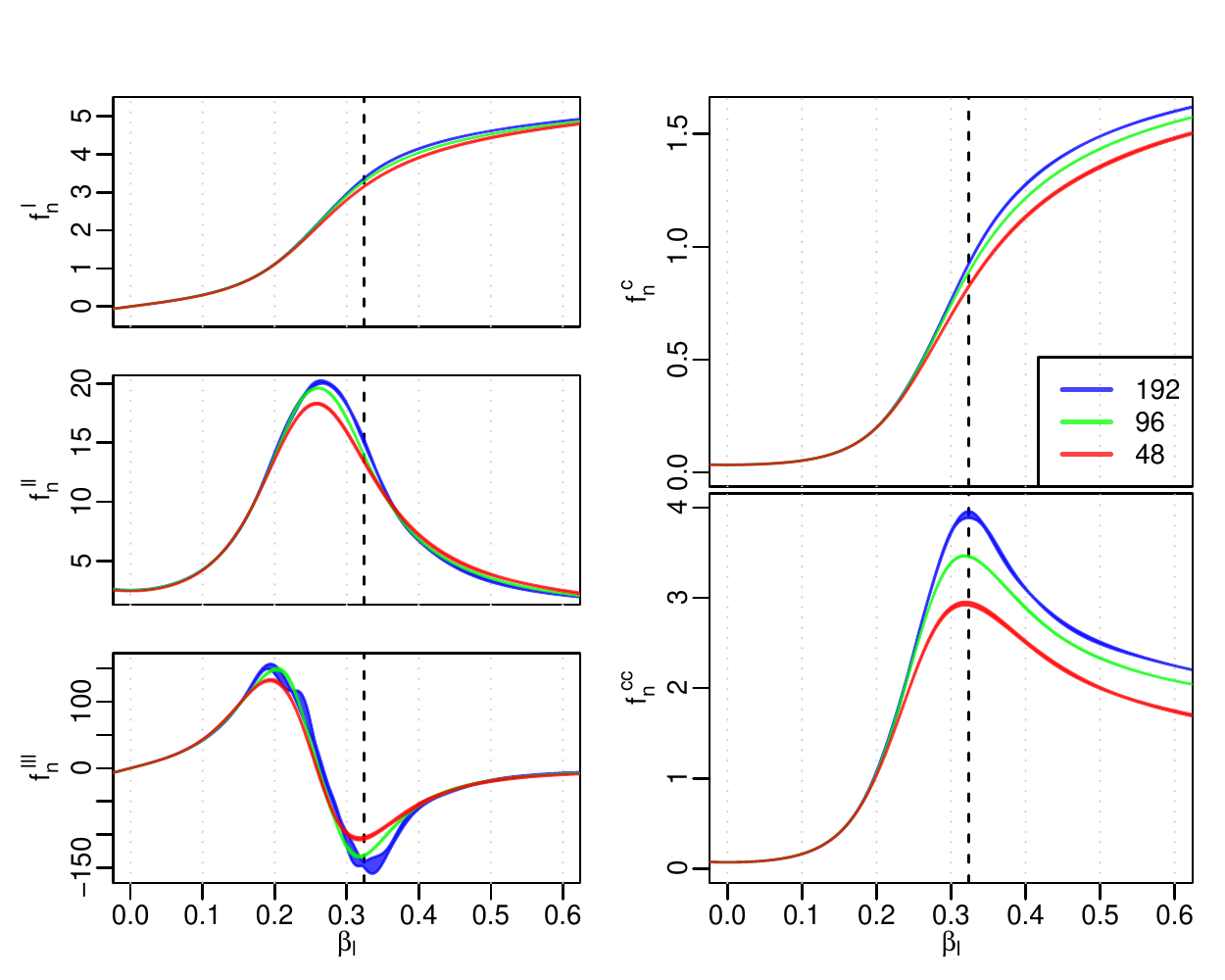}

\caption{\label{fig:scaling_l_c_bc=00003D0} The graphs on the left-hand side
depict the scaling of the estimates for first three derivatives of
the free energy with respect to $\beta_{l}$ and against $\beta_{l}$.
The two graphs on the right-hand side show the corresponding scaling
of the first two derivatives with respect to $\beta_{c}$. The couplings
are held fixed at $\left(\beta_{c}=0,\,\beta_{s}=1,\,\beta_{h}=0.5\right)$.
The vertical line on the left graphs marks the position ($\beta_{l}^{*}=0.247$)
of the peak in $f_{n=192}^{cc}$. We interpret the scaling of the
observables as a second-order stretched-writhed transition.}
\end{figure}

\begin{figure}[ht!]
\centering\includegraphics[width=10cm]{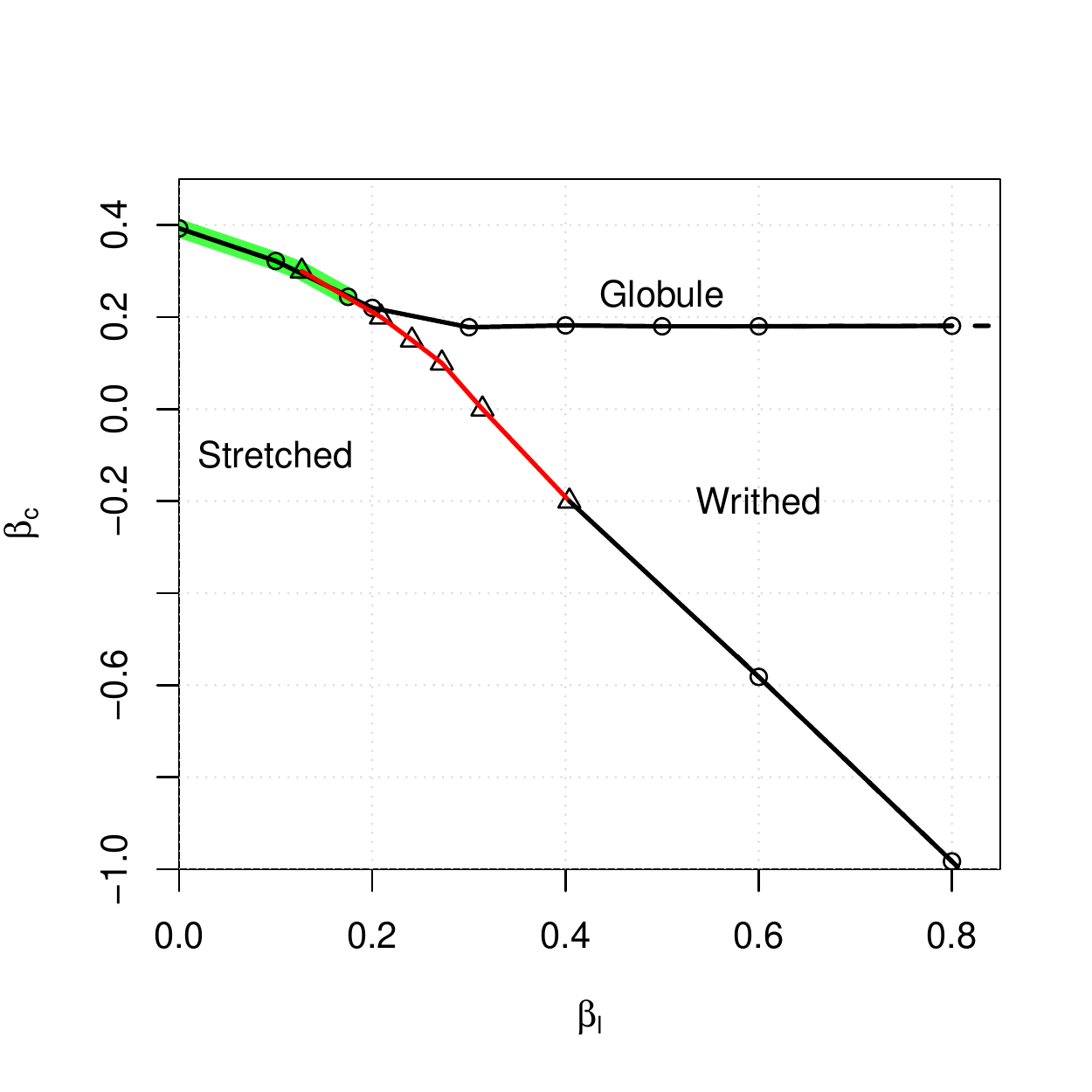}
\caption{\label{fig:phases_diagram_100_50} Finite size ($n=96$) phase diagram
at $\beta_{s}=1$, $\beta_{h}=0.5$. The black lines represent the
phase boundaries when determined from peaks in $f_{n}^{cc}$, the
red line when determined from the negative peaks in $f_{n}^{lll}$.
 The thick (green) line represents a line of confirmed
first-order transitions. The transition is second-order at $\beta_{l}=0.2$.}
\end{figure}

Figure~\ref{fig:scaling_l_c_bc=00003D0} shows the derivatives
at $\beta_{s}=1$ and $\beta_{h}=0.5$ against torque. In contrast
to the previous case (figure~\ref{fig:w_000}), we now detect scaling
in $f_{n}^{ll}$ and $f_{n}^{l\, l\, l}$, such that the negative
peak in the third derivative might diverge in the thermodynamic limit
(at least when the polymer is stiff enough). Other than the scaling
in $f_{n}^{cc}$, the scaling in $f_{n}^{ll}$ and $f_{n}^{l\, l\, l}$
is localized and restricted to the peak area. We find that the position
of the negative peak in $f_{n}^{l\, l\, l}$ coincides with the peak
in $f_{n}^{cc}$. This establishes further justification for having
identified the peak in $f_{n}^{cc}$ in Figure~\ref{fig:w_000} with
the position of the transition to produce the above phase diagram (\ref{fig:w_000-1}).
In order to produce the phase diagram for the stiff and pulled polymer,
we obtain the positions of the transitions either from the negative
peak in $f_{n}^{lll}\left(\beta_{l},\,\beta_{c}=\mbox{const}\right)$ or
the peaks in $f_{n}^{cc}\left(\beta_{l}=\mbox{const},\,\beta_{c}\right)$.
The resulting phase diagram is shown in Figure~\ref{fig:phases_diagram_100_50}.

\begin{figure}[ht!]
\centering\includegraphics[width=8cm]{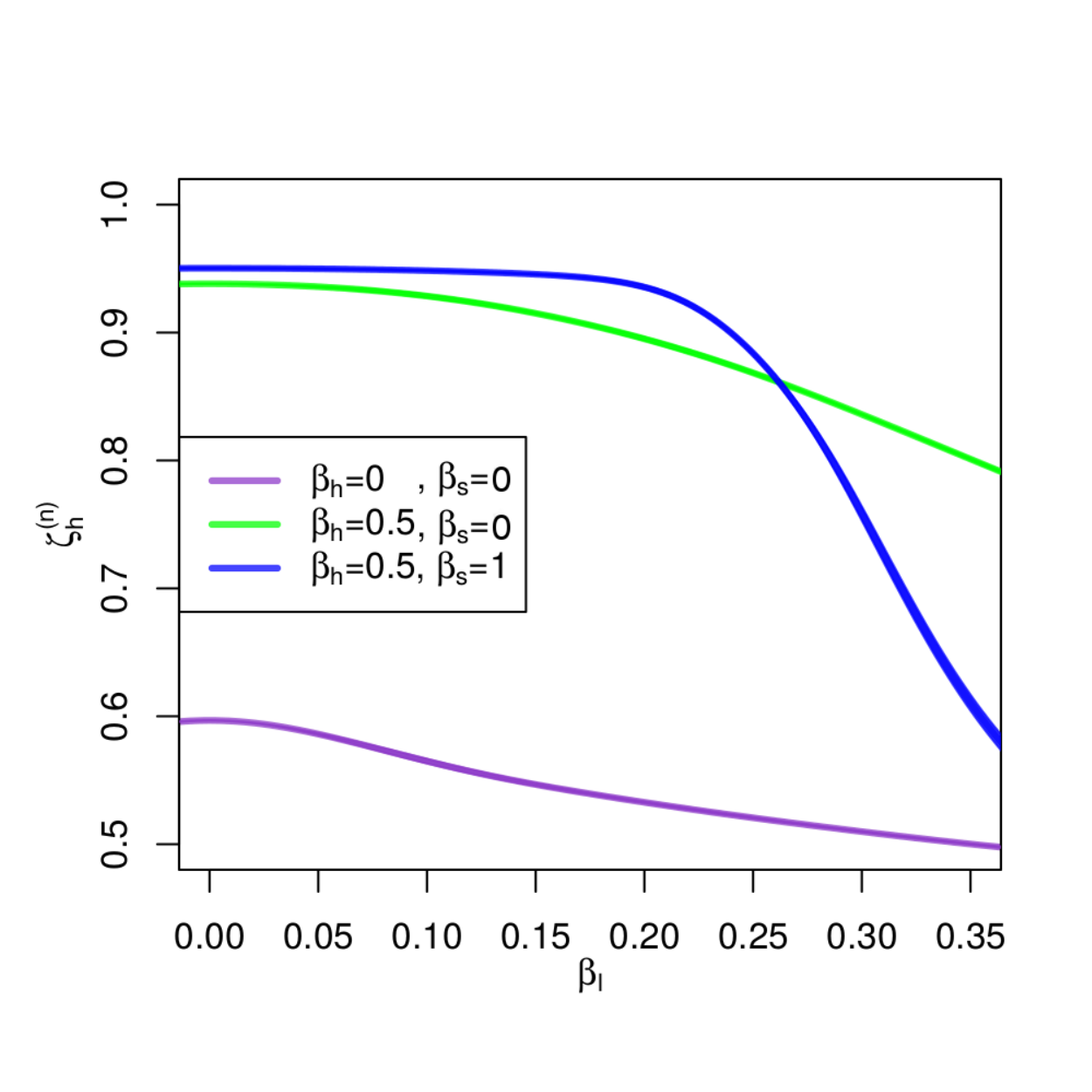}

\caption{\label{fig:PULLED_SAW_exponent_zeta_h} Estimate for $\zeta_{h}^{\left(n\right)}\left(\beta_{l}\right)$
at $n=48$ and different values of the pulling force: $\beta_h=0$ and $\beta_s=0$ (bottom), $\beta_h=0.5$ and $\beta_s=0$ (middle), and $\beta_h=0.5$ and $\beta_s=1$ (top). Changing to non-zero $\beta_h$ increases the exponent from values around $0.59$ to a value close to $1$ for small $\beta_l$, consistent with a stretched phase.}
\end{figure}

Due to the pulling, the coil-globule transition has become a first
order stretched-globule transition. The stretched phase is usually
characterized by a scaling exponent defined by $\left\langle h\right\rangle _{n}\sim n^{\zeta_{h}}$
where $\zeta_{h}=1$ as opposed to $\zeta_{h}=\nu_{SAW}\approx0.59$
when no pulling force is applied. This is shown in Figure~\ref{fig:PULLED_SAW_exponent_zeta_h}
which shows an estimate for 
\begin{equation}
\zeta_{h}^{\left(n\right)}\left(\beta_{l}\right):=\frac{\log\left(\left\langle h\right\rangle _{2n}/\left\langle h\right\rangle _{n}\right)}{\log\left(2\right)}
\end{equation}
at $n=48$ and $\beta_{h}=0,\,0.5$. Clearly, the result is compatible
with $\zeta_{h}=1$ at $\beta_{h}=0.5$. 

The phase diagram in Figure~\ref{fig:phases_diagram_100_50} is
associated with three transitions. The stretched-globule, stretched-writhed
and writhed-globule transition. We will examine these transitions
in the following.

\begin{figure}[ht!]
\centering\includegraphics[width=12cm]{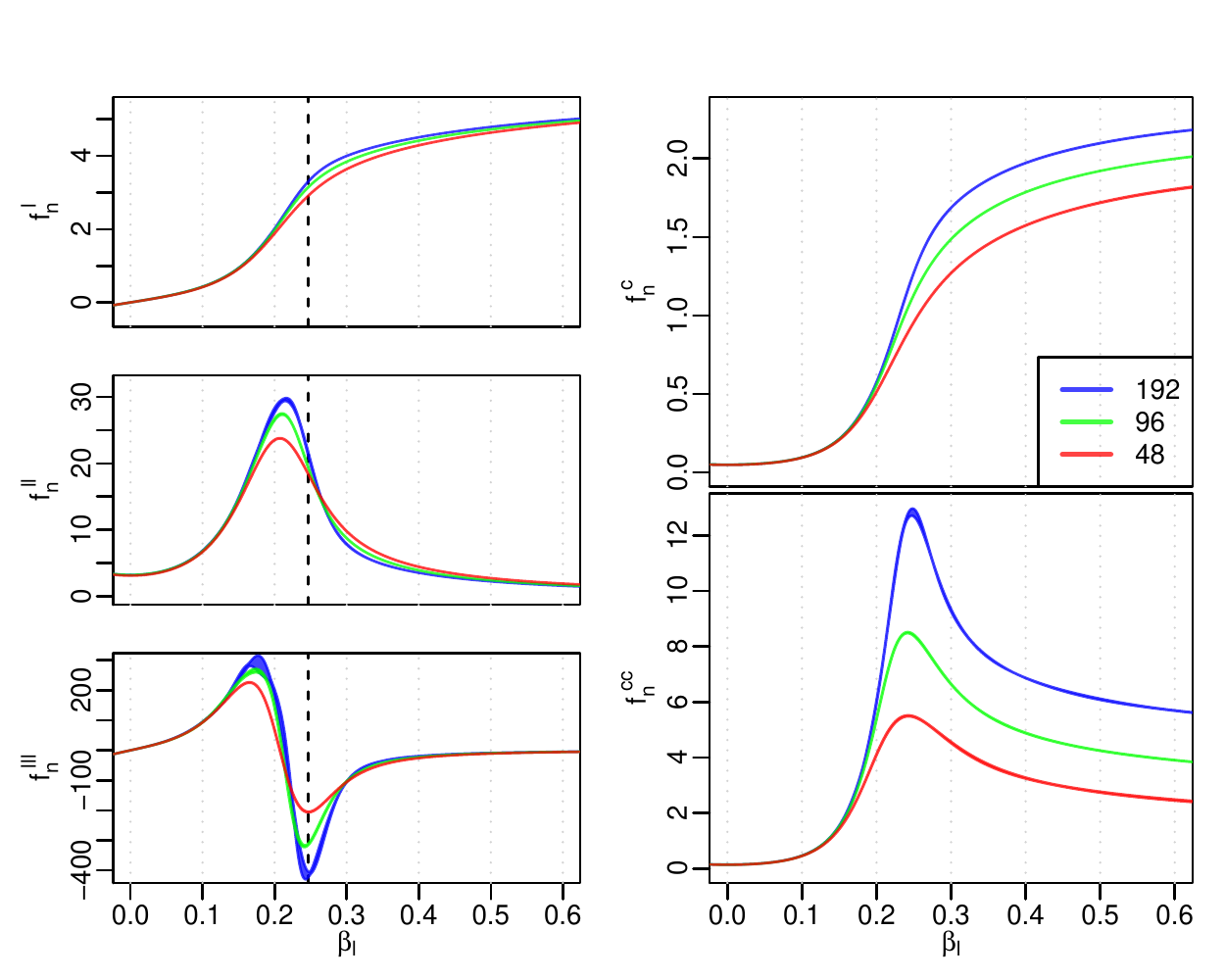}

\caption{\label{fig:scaling_l_c_bc=00003D0.15} The graphs on the left-hand
side depict the scaling of the estimates for first three derivatives
of the free energy with respect to $\beta_{l}$ and against $\beta_{l}$.
The two graphs on the right-hand show the corresponding scaling of
the first two derivatives with respect to $\beta_{c}$. The couplings
are held fixed at $\left(\beta_{c}=0.15,\,\beta_{s}=1,\,\beta_{h}=0.5\right)$.
The vertical line on the left graphs marks the position ($\beta_{l}^{*}=0.247$)
of the peak in $f_{n=192}^{cc}$. We interpret the scaling of the
observables as a second-order stretched-writhed transition.}
\end{figure}

\begin{figure}[ht!]
\centering\includegraphics[width=10cm]{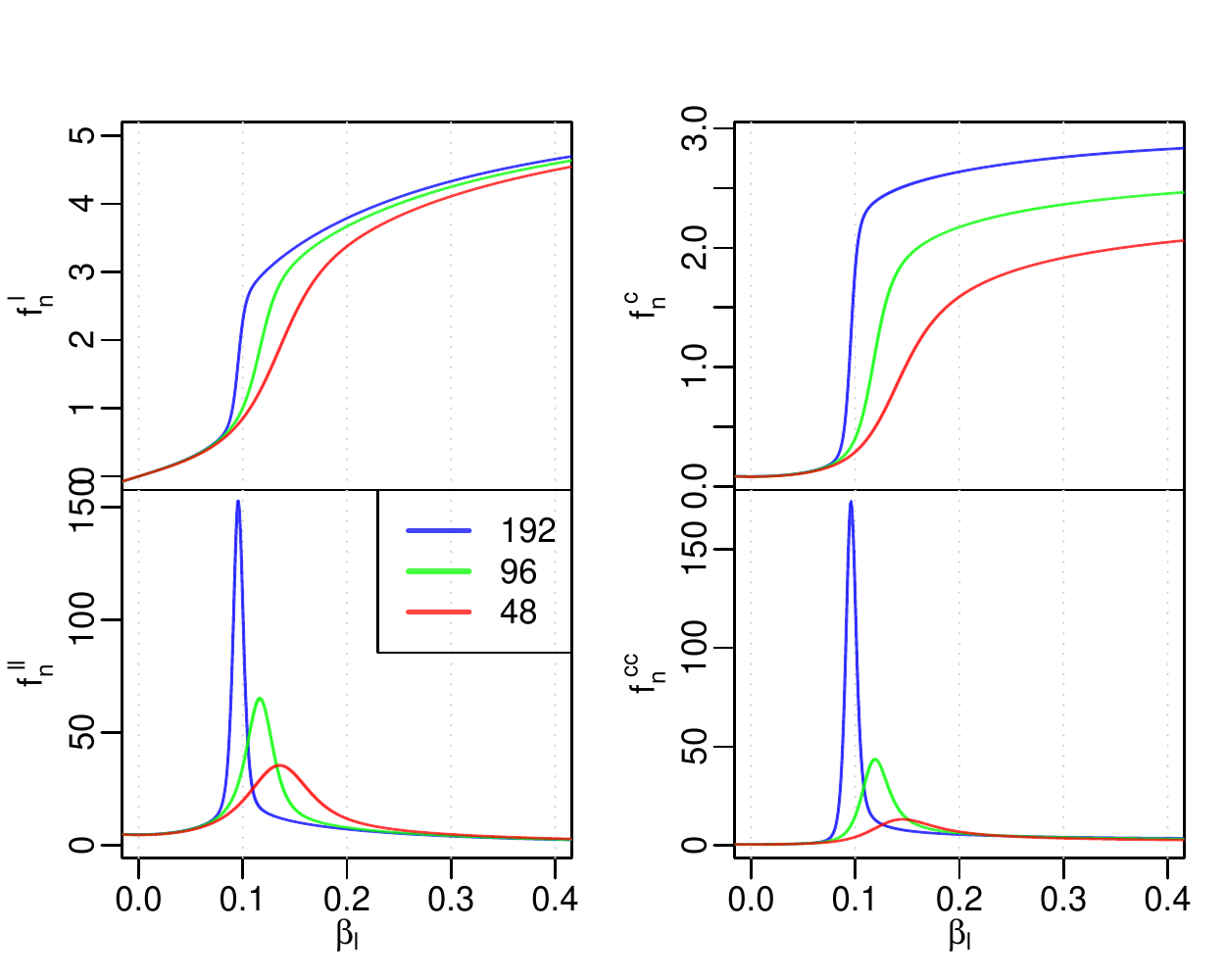}

\caption{\label{fig:scaling_l_c_bc=00003D0.3} The graphs on the left-hand
side depict the scaling of the estimates for first two derivatives
of the free energy with respect to $\beta_{l}$ and against $\beta_{l}$.
The two graphs on the right-hand show the corresponding scaling of
the first two derivatives with respect to $\beta_{c}$. The couplings
are held fixed at $\left(\beta_{c}=0.30,\,\beta_{s}=1,\,\beta_{h}=0.5\right)$.
We interpret the scaling of the observables as the first-order collapse
transition (stretched-globule transition).}
\end{figure}

In Figures~\ref{fig:scaling_l_c_bc=00003D0}, \ref{fig:scaling_l_c_bc=00003D0.15}, and \ref{fig:scaling_l_c_bc=00003D0.3},
we show transitions out of the stretched phase by applying torque at different
values of $\beta_{c}$. We have already discussed the case $\beta_{c}=0$
in Figure~\ref{fig:scaling_l_c_bc=00003D0}, where we pass through the stretched-writhed
transition. At $\beta_{c}=0.3$, we pass through the stretched-globule transition.
The derivatives of the free energy are shown in Figure~\ref{fig:scaling_l_c_bc=00003D0.3}.
We point out that not only does the number of contacts behave like a discontinuous
order parameter, but also writhe fluctuations per length appear to
scale linearly with the length. This of course confirms that the stretched-globule
transition remains first-order when some torque is applied. Figure~\ref{fig:scaling_l_c_bc=00003D0.15}
considers the transition near the triple point. By setting $\beta_{c}=0.15$
we begin in the stretched phase at low torque and end up in the writhed
phase at high torque. The scaling is compatible with a second-order
phase transition. It is clearly more pronounced than at $\beta_{c}=0$. 

\begin{figure}[ht!]
\centering\includegraphics[width=8cm]{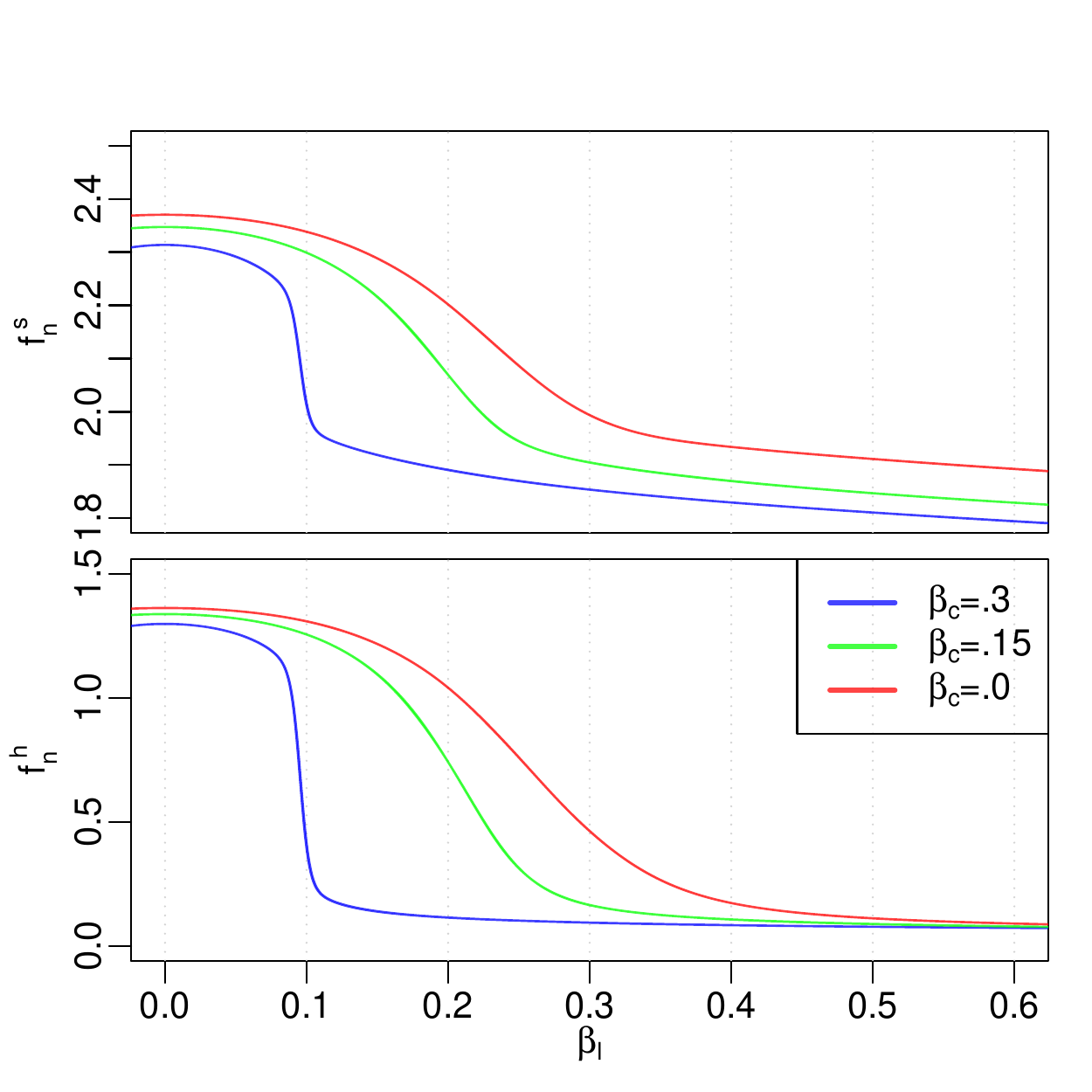}

\caption{\label{fig:expectations} Expectation values per length of the number
of stiff sites $f_{n}^{s}$ and extension $f_{n}^{h}$ at $n=192$
and values $\beta_{c}=0,\,0.15,\,0.3$, where the upper curves in
each graph corresponds to $\beta_{c}=0$ and the lowest curves corresponds
to $\beta_{c}=0.3$.}
\end{figure}

Figure~\ref{fig:expectations} shows the behavior of the number of
stiff sites and the extension across the transitions. Both observables
decrease across the transitions. 
For $\beta_c=0$ we find a weak second-order stretched-writhed transition, whereas for $\beta_c=0.3$ we find a strong first-order stretched-globule transition, confirming
the conclusions obtained from the scaling of the free energy.

\section{Conclusion}

In this paper, we considered an ensemble of geometrically and topologically
constrained SAWs on the face-centered cubic lattice. We weighted these
SAWs by their writhe and interpreted this as a model of twist storing
polymers like DNA. In this framework, the coupling of the writhe is
regarded as being related to a torque applied to the DNA. In addition,
we also weighted the number of contacts in the SAW by a coupling related
to the quality of the solvent. Via simulations, we examined the finite
size system for signatures of phase transitions in the coupling space
spanned by torque and solvent quality. Our results suggest that weighting
the writhe of flexible SAWs induces a second-order phase transition
from a coiled phase into a writhed phase. However, this transition
is very weak and at the lengths considered here, we were not able to observe
it via the moments of the writhe. However, the number of contacts shows weak scaling
so that by considering the problem in the torque-solvent quality plane,
we were able to conjecture that the transition should lie at $\beta_{l}\approx0.3$.
In this plane, we conjecture three phases exist: coil or extended, globule and writhed. The first two of these are well known from standard polymer collapse. It will be interesting in the future to investigate more closely the transition into the writhed phase as we were unable to estimate critical exponents. The nature of the point between the three phases would also be of interest.

Then, we considered the case of pulled, semi-flexible polymers. When pulling a SAW the coil (extended) phase immediately becomes a stretched phase with anisotropic size scaling. However, both the globule phase, as has previously been ascertained, and the writhed phase remain. The addition of pulling seems to  make the stretched-writhed and stretched-globule transitions more pronounced than their coil-writhed and coil-globule counterparts. In particular, the stretched-globule transition becomes first-order, as is well known. The writhed-globule and stretched-writhed transitions seem to be second-order although our data is once again not sufficient to meaningfully estimate the associated scaling exponents.

\ack

One of the authors, ED, gratefully acknowledges the financial support
of the University of Melbourne via its Melbourne International Research
Scholarships scheme. Financial support from the Australian Research
Council via its support for the Centre of Excellence for Mathematics
and Statistics of Complex Systems and the Discovery Projects scheme (DP160103562)
is gratefully acknowledged by one of the authors, ALO.

\section*{References}


\end{document}